# The Measurement Process in Relational Quantum Mechanics

B.K. Jennings

Understanding the quantum measurement problem is closely associated with understanding wave function collapse. Motivated by Breuer's claim that it is impossible for an observer to distinguish all states of a system in which it is contained, wave function collapse is tied to self observation in the Schmidt biorthonormal decomposition of entangled systems. This approach provides quantum mechanics in general and relational quantum mechanics in particular with a clean, well motivated explanation of the measurement process and wave function collapse.

Relational quantum mechanics[1] provides a parsimonious interpretation of quantum mechanics that has many compelling features. However the nature of the measurement process and wave function collapse is left rather vague. In this paper their nature is clarified using the Schmidt biorthonormal decomposition and self observation. While the arguments are presented in the frame work of relational quantum mechanics they are quite general, depending only on the quantum state being described by a probability amplitude.

Any entangled state of quantum systems A and B can be expanded in a Schmidt biorthonormal decomposition $\Psi(A, B) = \sum_i c(i) |A(i)>|B(i)>$, where the $|A(i)>$ and $|B(i)>$ are orthonormal bases in the two spaces respectively[2]. The $c(i)$ are complex coefficients. The decomposition is unique except when two or more coefficients have equal magnitudes as, for example, when two spin-a-half particles are coupled to spin zero. Then there can be any linear combination in the degenerate space. There is also a trivial ambiguity about how the phase of a single term is distributed between the two states and the expansion coefficient. This has no observable or, indeed, any meaningful consequences.

Thomas Breuer showed[3] that neither A nor B can measure the phases of the $c(i)$ in the Schmidt biorthonormal decomposition of their entangled state if the union of A and B is an isolated system. This results leads to an heuristic argument for when the wave function collapses. If there are three entangled observers they would disagree on which phases are unmeasurable corresponding to the three different partitions of the three observer system into two systems. Therefore they will each ascribe a different wave function to the three observer system. This is only consistent if the wave function is relative not absolute. Now relative to A or B, the entangled state can not be due to unitary time evolution with a known interaction, as it would be relative to a third party, since if it was the unknowable phases could be determined. Similarly both A and B must observe that they are in a pure state corresponding to one term in the Schmidt biorthonormal decomposition since only such a pure state does not depend on the unknowable phases. Secondly, since the wave function is relative, suggesting it is epistemic, self observation would naturally result in a system knowing which Schmidt biorthonormal decomposition state it is in producing wave function collapse. Relative to itself a system is always in a definite state in a given Schmidt biorthonormal decomposition.

The conclusion of the above arguments is that the collapse of the wave function is caused by self observation of the systems in an entangled pair: $\Psi(A,B) = \sum_i c(i) |A(i)>|B(i)> \implies |A(j)>$

$|B(j)>$ for a given $j$ i.e. the collapsed wave function is $\Psi(A,B) = |A(j)>|B(j)>$. The collapse is relative to the systems A and B and always occurs when two systems become entangled but would not apply relative to a system not interacting with A or B. With this motivation and definition, self observation collapsing the wave function can be added as a separate axiom to relational quantum mechanics giving a modified or augmented relational quantum mechanics. This leads to no contradictions or internal inconsistencies but a clear description of the measurement process. It would lead to contradictions if the wave function were absolute not relative, but Breuer's result implies it can not be absolute.

It is worth noting that decoherence[4] by itself cannot cause[5] wave function collapse. Decoherence only makes the reduced density matrix diagonal, diagonal in the basis defined by the Schmidt biorthonormal decomposition e.g. $\rho = \sum_i |B(i)>|c(i)|^2<B(i)|$. This can be easily seen by taking the partial trace of an entangled pair using the Schmidt biorthonormal decomposition. Self observation on the other hand collapses that reduced density matrix to a single term, $\rho = |B(j)><B(j)|$. Due to the Schmidt biorthonormal decomposition, both decoherence and self observation are with respect to the same basis states. This makes it difficult to disentangle the effects of the two very different processes. To the extent one has contra-factual definiteness it is impossible to determine phenomenologically when wave function collapse occurs.

A measurement of an observed system, A, with respect to an observer system, B, has three aspects. First the two system must interact to generate an entangled final state. Second the initial values of the property being measured must be correlated with the states in the Schmidt decomposition of the final state. And third, self observation must cause the collapse of the entangled final state to give a definite result for the measurement. Since the initial state of the observer, B, can be assumed to be known the relevant part of the unitary transformation generated by the interaction can be written $U = \sum_i |A'(i)>|B'(i)><B(init)|<A(i)|$ yielding the collapsing wave function $\Psi(A,B) = \sum_i <A(i)|A(init)>|A'(i)>|B'(i)> \Rightarrow |A'(j)>|B'(j)>$ for fixed $j$. The $A(i)$ are the eigenvectors of the Hermitian operator for the property being measured and the relation $A'(i) = A(i)$ is frequently assumed but is not necessary. The sets $A(i)$, $A'(i)$, and $B'(i)$ are orthonormal in their respective spaces. Thus $\Psi(A,B) = \sum_i <A(i)|A(init)>|A'(i)>|B'(i)>$ is the Schmidt biorthonormal decomposition of the entangled state. From ref. 1, sec. III C, the probability of collapse to a given $j$ is $|<A(j)|A(init)>|^2$. The probability is frequentist not Bayesian i.e. quantum mechanics predicts the ratio of different outcomes if the experiment is repeated many times.

Some comments are in order. The need for the first point is self evident; for a measurement to be performed the systems must interact. However the interaction does not need to be direct but can be by way of one or several intermediate systems. All that is required is that the second point be valid. The second point is required so that the collapse to a given state in the Schmidt biorthonormal decomposition gives definite information on a property of the initial state. Regarding the third point, a measurement is only complete when the wave function has collapsed by self observation; it is only then that one has a measured result or a relative fact, namely the value of $j$ (and related information) in the collapsed wave function. Wave function collapse also reduces the measurement to a series of yes-no questions: Is the state $j$, for each value of $j$, occupied? This is very much inline with the discussion of yes-no questions in the original discussion of relational quantum mechanics[1].

Two predictions follow from the above considerations. First, since the wave function is relative there can be no physical effect, such as radiation emission[6], generated from wave function collapse since the collapse can occur at different times and places (or not at all) for different observers. Second, since decoherence is not an essential part of the measurement process, an experimentalist can perform a measurement on a quantum system with no decoherence taking place between the quantum system and the experimentalist.

In conclusion, wave functions are relative not absolute, wave function collapse is due to self observation, and a quantum measurement occurs when two quantum systems becoming entangled. The resulting entangled wave function collapsing by self observation then yields the definite result of the measurement. The collapse is into one of the terms in the Schmidt biorthonormal decomposition of the entangled state.